\documentclass[reprint, amsmath, amssymb, aps, prx,longbibliography,superscriptaddress,floatfix]{revtex4-1}
\usepackage{amsmath,amssymb,bm,graphicx}
\usepackage{graphics}
\usepackage{float} 
\usepackage[stable]{footmisc}
\usepackage{soul}
\usepackage[breaklinks=true,colorlinks,citecolor=blue,linkcolor=blue,urlcolor=blue]{hyperref}
\bibpunct{[}{]}{,}{n}{}{}

\def\be{\begin{equation}}
\def\ee{\end{equation}}
\def \bea{\begin{eqnarray}}
\def \eea{\end{eqnarray}}

\usepackage[colorlinks, linkcolor= blue, citecolor = blue, urlcolor=blue]{hyperref}

\def \be {\begin{equation}}
\def \ee {\end{equation}}
\def \bea{\begin{eqnarray}}
\def \eea{\end{eqnarray}}

\def \formula{$\mathrm{MoSi}_{2}{\mathrm{Z}}_{4}$}
\def \formulaAs{$\mathrm{MoSi}_{2}{\mathrm{As}}_{4}$}
\def \formulaN{$\mathrm{MoSi}_{2}{\mathrm{N}}_{4}$}
\def \formulaP{$\mathrm{MoSi}_{2}{\mathrm{P}}_{4}$}
\def \hsp{\hspace{5cm}}

\begin{document}

\title{Strongly bound excitons in monolayer MoSi$_2$Z$_4$ (Z = pnictogen)}
\author{Pushpendra Yadav}
\email{pyadav@iitk.ac.in}
\affiliation{Department of Physics, Indian Institute of Technology Kanpur, Kanpur-208016, India}
\author{Bramhachari Khamari}
\affiliation{Department of Physics, Indian Institute of Technology Kanpur, Kanpur-208016, India}
\author{Bahadur Singh}
\affiliation{Department of Condensed Matter Physics and Materials Science, Tata Institute of Fundamental Research, Mumbai 400005, India}
\author{K. V. Adarsh}
\affiliation{Department of Physics, Indian Institute of Science Education and Research Bhopal, Bhopal 462066, India}
\author{Amit Agarwal}
\email{amitag@iitk.ac.in}
\affiliation{Department of Physics, Indian Institute of Technology Kanpur, Kanpur-208016, India}

\begin{abstract}Reduced dielectric screening in two-dimensional materials enables bound excitons, which modifies their optical absorption and optoelectronic response even at room temperature. Here, we demonstrate the existence of excitons in the bandgap of the monolayer family of the newly discovered synthetic {\formula} (Z = N, P, and As) series of materials. All three monolayers support several bright and strongly bound excitons with binding energies varying from 1 eV to 1.35 eV for the lowest energy exciton resonances. On increasing the pump fluence, the exciton binding energies get renormalized, leading to a redshift-blueshift crossover. Our study shows that the {\formula} series of monolayers offer an exciting test-bed for exploring the physics of strongly bound excitons and their non-equilibrium dynamics.
\end{abstract}
\maketitle
\section{INTRODUCTION}
\label{INTRO}
The interplay of light-matter interactions and Coulomb interactions in two-dimensional (2D) materials gives rise to  very exciting physics~\cite{Mueller2018,Thygesen_2017,Wang-excitonsTMD-colloquium,Mounet2018,amit-nanoscale-2018}. 
Compared to bulk materials, layered 2D materials possess many peculiar properties strongly related to their number of layers. 
Coupled with the rapidly growing family of stable 2D materials, this has accelerated the exploration of novel optical responses and their potential use for next-generation  optoelectronic
device applications~\cite{Novoselov2004_graphene1,Bernardi2013-graphene-light,Butler2013}.
 In particular,  2D monolayers of transition metal di-chalcogenides (TMDs) are known to exhibit prominent excitonic effects~\cite{optical-MoS2-PRL,strain-exciton-Mos2-PRB,Dark-Exciton-PhysRevMaterials.2.014002,exciton-TMD}, due to the reduced dielectric screening and enhanced Coulomb interactions. These excitonic effects modify the optical absorption spectrum significantly and can be tuned by a variety of external stimuli, opening up new possibilities for optoelectronic applications~\cite{Butler2013,Bernardi2013-graphene-light}. 

Recently, a new family of materials (MoSi$_2$N$_4$ class of materials) has been added to the family of 2D monolayers. These are synthetic materials, with no known naturally occurring three-dimensional counterparts~\cite{Hong670}. 
In contrast to the monolayers of naturally occurring crystals, these synthetic 2D materials were designed using a bottom-up approach, and a monolayer of MoSi$_2$N$_4$ was grown using chemical vapour deposition. Additionally, several other materials of the same family of compounds, such as {\formulaAs}, $\mathrm{WSi}_{2}{\mathrm{N}}_{4}$, $\mathrm{WSi}_{2}{\mathrm{V}}_{4}$, were predicted to be dynamically stable~\cite{Hong670}. 
The {\formula} series of materials has been shown to have interesting electrical~\cite{Wu2021-APL,Cao2021-APL,Guo_2020-IOP,Bafekry2020MoSi2N4SA,keshari}, thermal~\cite{Yu_2021}, optical~\cite{Yang2021-non-linear-optical-response,Yao2021-Nanomaterials,SHG-PRB}, valley~\cite{Valley-1,Yang_valley-valley2,ai2021theoretical-valley3}, and spin dependent~\cite{Rajibul-Barun-spin} properties. However, the physics of excitons in this series of materials is relatively less explored~\cite{exciton-1-wu2021mosi2n4}. 

In this paper, we investigate the equilibrium and non-equilibrium optical properties of monolayer {\formula} series of compounds (with Z $=$ N, As and P), focusing on the excitonic effects.
To obtain accurate electronic properties, we use the density functional theory (DFT) calculations and include the quasiparticle (QP) self-energy corrections using quantum many-body perturbation theory (MBPT) following the GW method~\cite{Hedin-GW1,Hybertsen-GW2,DFT_GWA_Exciton-GW3}. We show that the monolayer {\formulaAs} and {\formulaP} have a direct QP bandgap of 1.70 and 1.74 eV respectively, while monolayer {\formulaN} hosts an indirect QP bandgap of 3.58 eV with a comparable direct bandgap~\cite{Hong670}. 
\begin{figure*}
	\includegraphics[width=0.9\textwidth]{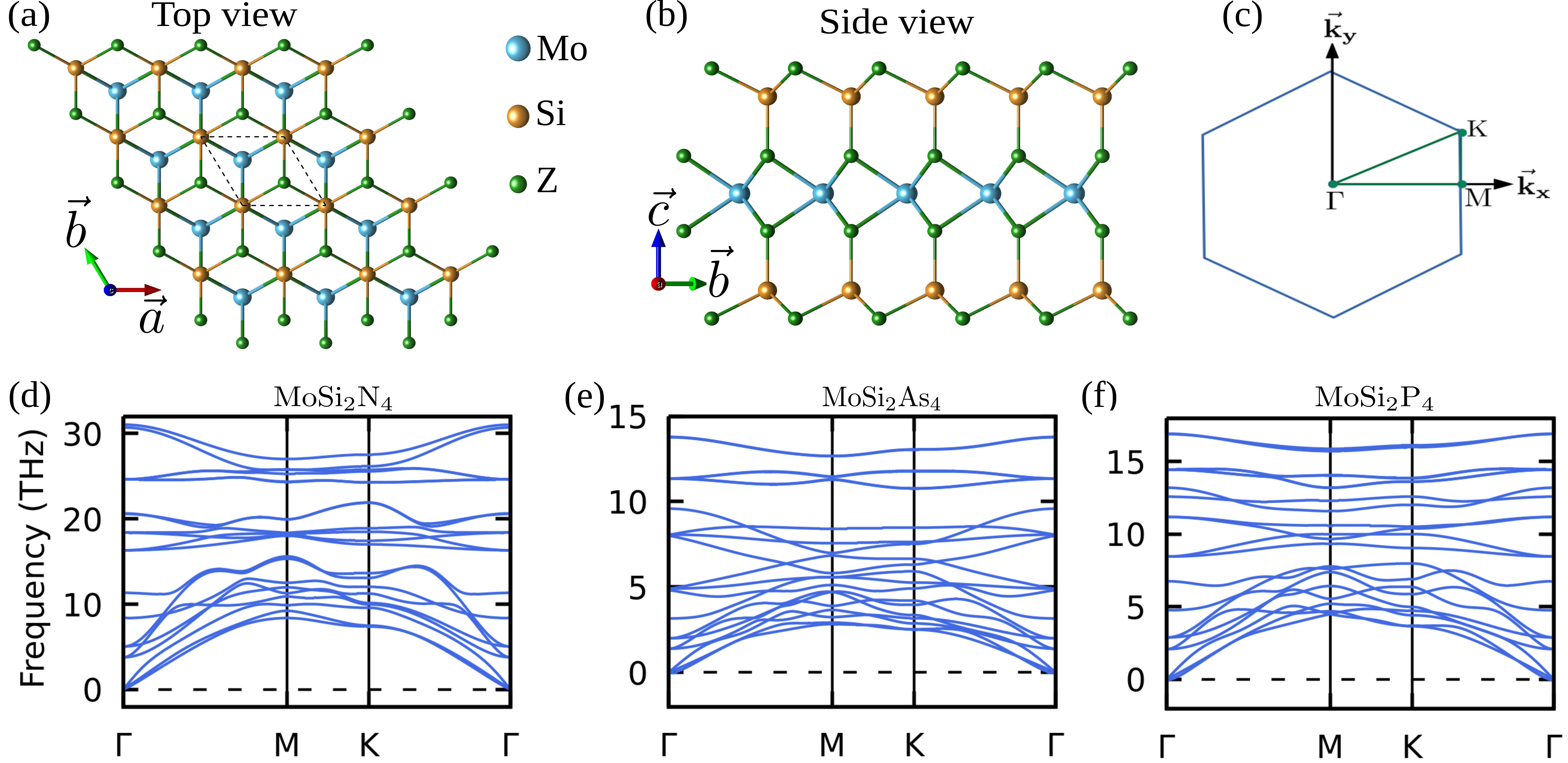}
	\caption {(a) The top and (b) side view of the monolayer crystal structure of {\formula} (Z= N, As or P). The  Z-Si-Z-Mo-Z-Si-Z arrangement of atoms along the $c$-axis can be clearly seen in panel (b). (c) The 2D hexagonal Brillouin zone (BZ). (d)-(e) The phonon dispersion of the {\formulaN}, {\formulaAs}, and {\formulaP} monolayer respectively. All three of these {\formula} series monolayers display no negative frequency over the entire BZ and are mechanically stable.
	\label{str}}
\end{figure*}

To study the excitonic resonances and their impact on the optical absorption spectrum, we include electron-hole correlations on top of the GW ground state, using the Bethe-Salpeter equation (BSE)~\cite{BSE-1-Strinati,BSE-2-Strinati,BSE-3-Onida}. Our calculations show that in contrast to two excitonic peaks in the QP gap region found in MoS$_2$~\cite{optical-MoS2-PRL,spin-orbit-MoS2-PRB,amit-exciton-1,amit-exciton-2}, the {\formula} series of materials host three or more strongly bound bright excitonic peaks in the band-gap region~\cite{exciton-1-wu2021mosi2n4}. Compared to other 2D materials, the lowest energy exciton peak in all three monolayers has a very high binding energy (BE) of 1 eV or more. 
We also explore the impact of pump fluence on the renormalization of the exciton BE, within the framework of the  time-dependent Bethe-Salpeter equation (td-BSE). We find that on increasing the pump fluence, the exciton BE initially decreases, owing to the additional screening arising from the photo-excited charge carriers. However, with increasing the pump fluence, the photo-excited carriers and the exciton density increases significantly,  and exciton-exciton repulsion starts dominating. This results in a redshift-blueshift crossover in the exciton BE~\cite{amit-exciton-2}, on varying the pump fluence. 
Our study establishes the {\formula} series of compounds as an interesting platform for i) exploring the physics of strongly bound exciton in 2D materials and ii) 
exploring optoelectronic applications in the infrared (MoSi$_2$As$_4$ and MoSi$_2$P$_4$ with an optical bandgap of $\sim 0.7$ eV) and visible regime (MoSi$_2$N$_4$ with an optical bandgap of 2.35 eV).

\begin{figure*}[t]
	\centering
	\includegraphics[width =0.9\textwidth]{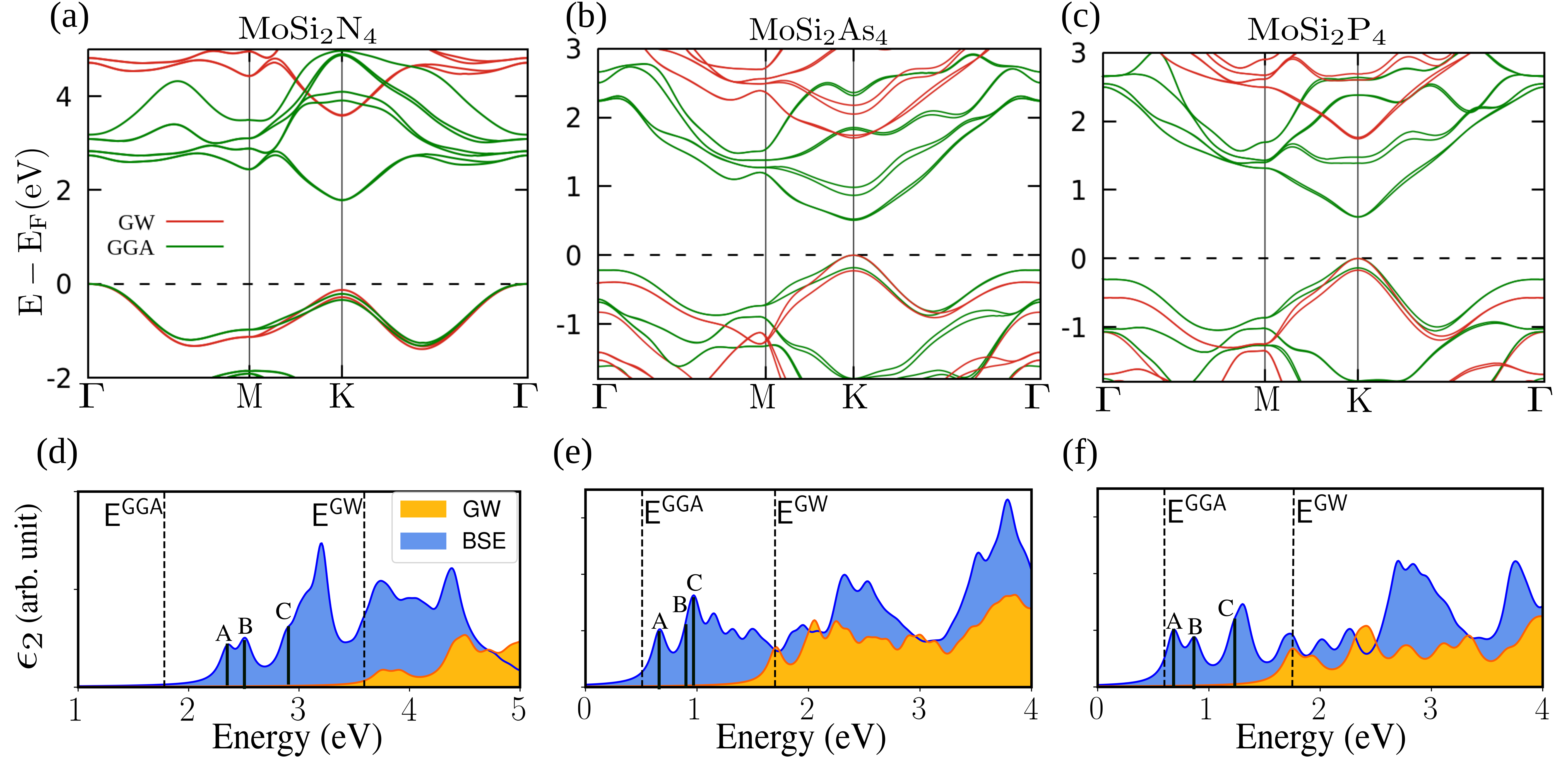}
	\caption {The electronic bandstructure calculated in presence of spin-orbit coupling (SOC) using GGA (green color) and the QP  bandstructure calculated using evGW (red color) for monolayer (a) {\formulaN}, (b) {\formulaAs}, and (c) {\formulaP}. The corresponding optical absorption spectra, calculated using the Bethe-Salpeter equation on top of the GW bandstructure (with SOC), for the three monolayers are shown in (d), (e), and (f), respectively. The bandgap calculated within the GGA approximation and the GW scheme are marked by dashed vertical lines. The BSE optical spectrum includes the two-particle electron-hole interactions (EHI) and captures the excitonic resonances, which manifest as several prominent absorption peaks below the GW bandgap. The location of the first three prominent bright exciton peaks, similar to the A, B, and C peaks in monolayer MoS$_2$, is marked by black lines for all three synthetic monolayers. In contrast to monolayer MoS$_2$, which hosts two prominent exciton peaks in the bandgap region, the {\formula} monolayers host several (more than three) bright exciton peaks in the electronic bandgap.}
	\label{BS}
\end{figure*}

This paper is organized as follows. In Sec.~\ref{CSCM} we discuss the crystal structure and details of the computational methodology. The electronic properties, the optical absorption spectrum, and the exciton binding energies and wave function for prominent bright excitons are presented in Sec.~\ref{EQP}. In Sec.~\ref{Non-EQL}, we explore the fluence dependent non-equilibrium optical properties such as photo-excited charge carrier dependent screening effects leading to shifting of the excitonic peaks and binding energies. We summarize our findings in Sec.~\ref{SC}.

\section{CRYSTAL STRUCTURE AND COMPUTATIONAL METHODS}
\label{CSCM}
The monolayer {\formula} has a hexagonal lattice structure 
[see Fig.~\ref{str} (a) and (b)], which breaks the space inversion symmetry~\cite{Hong670,Valley-1}. For our $ab-initio$ calculations we have used the previously reported lattice parameters $a=2.909$ {\AA} for {\formulaN}, $a=3.621$ {\AA} for {\formulaAs}, and $a=3.471$ {\AA} for {\formulaP} by Hong {\it et. al.} in Ref.~\cite{Hong670}, and relaxed the structure until the residual force on each atom becomes less than 0.0001 eV/{\AA}. We add a $27$ {\AA} vacuum along the out of plane axis, to avoid spurious inter-layer interactions.

To confirm the dynamical stability of the monolayer {\formula} structure, we have performed the phonon calculations with a $2\times 2\times 1$ supercell. For this, we have used the first-principle calculations based on density functional theory (DFT) as implemented in the Vienna ab-initio simulation package (VASP)~\cite{PhysRevB.54.11169,PhysRevB.59.1758}. The exchange-correlation effects are treated within the generalized gradient approximation (GGA)~\cite{GGA-1,GGA-PBE}. 
An energy cutoff of 500 eV for the plane-wave basis set and tolerance of $10^{-7} $ eV is used for electronic energy minimization. After calculating the ground state charge density self-consistently, we use the Phonopy code package~\cite{phonon} to get the phonon dispersion of all three {\formula} monolayers. The phonon dispersion of the monolayer {\formulaN}, {\formulaAs}, and {\formulaP} are shown in Fig.~\ref{str} (d)-(f). We find that all three of these {\formula} series monolayers display no negative frequency over the entire Brillouin zone (BZ) and are mechanically stable.

As an additional check for the GGA bandstructure, we use the relaxed structure and validate our DFT calculations using the Quantum ESPRESSO package~\cite{QE}   
with fully relativistic norm-conserving pseudo-potential. 
To perform the BZ integration, we used a $\Gamma$-centered $12\times 12 \times 1$ Monkhorst $k$ mesh~\cite{PhysRevB.13.5188}.
To simulate the QP energy and the optical excitation calculations, we have used quantum MBPT following the implementation in the YAMBO package~\cite{yambo20091392,yambo2019}. The ground-state Kohn-Sham~\cite{Kohn-Sham} eigenfunctions obtained from the Quantum ESPRESSO based GGA calculations are used as the initial input by the MBPT. 

\begin{table*}[t]
\caption{The electronic bandgap for the three {\formula} monolayers with spin-orbit coupling (SOC), calculated within the GGA approximation ($\mathrm{E_{g}^{GGA}}$), with hybrid (HSE) functional ($\mathrm{E_g^{HSE}}$)~\cite{HSE06-1,HSE06-2}, the QP GW bandgap ($\mathrm{E_g^{GW}}$), the optical bandgap, and the direct/indirect nature of the bandgap. All three of these synthetic monolayers support very strongly bound excitons with binding energies of 1 eV or more, which are listed in Table~\ref{table:2}.}	
\vspace{0.15 cm}
\begin{tabular}{c c c c c c}

	\hline \hline \vspace{1 mm}
	Structure \hsp & $\mathrm{E_g^{GGA}}$(eV) \hsp & $\mathrm{E_g^{HSE}}$ (eV)~\cite{Yang2021-non-linear-optical-response}\hsp&  	$\mathrm{E_g^{GW}}$ (eV) \hsp   &Optical bandgap (eV)\hsp  &   Nature of bandgap \\
	\hline
	{\formulaN} \hsp & 1.78 \hsp& 2.24 \hsp & 3.58  \hsp & 2.35 \hsp& Indirect\\
	{\formulaAs} \hsp & 0.51 \hsp& 0.88 \hsp & 1.70 \hsp & 0.66 \hsp& Direct\\
	{\formulaP} \hsp& 0.60 \hsp& 1.01 \hsp & 1.74  \hsp & 0.68 \hsp& Direct\\
	\hline
\end{tabular}
\label{table:1}
\end{table*}

 For incorporating the QP self-energy corrections in the electronic structure, we have used the GW method as implemented in the YAMBO package~\cite{yambo20091392,yambo2019}. To evaluate the diagonal elements of exchange self-energy (to evaluate the Coulomb integrals), we have used $10^6$ random points in our calculation with a $G$-vector
cutoff of 14 Ry. This integral has been evaluated using a Monte Carlo scheme, known as the random integration method~\cite{Pulci1998-RIM,Rozzi2006-Coulomb-truncation}. The numerical integral has been defined within a box-type geometry of size $50.01$ {\AA} on either side of the monolayer \formulaN. To calculate the polarization function within the random-phase approximation, we have used three hundred forty bands and a response block size of $15$ Ry after a convergence test. Following this, we used a plasmon-pole approximation~\cite{PPA} to calculate the inverse of the microscopic dynamic dielectric function. A self-consistent GW on eigenvalues only (evGW) is adopted for the QP self-energy calculations.
Using the Kohn-Sham wave functions and the calculated QP energies, we have calculated the optical-spectra using the BSE. The linear response optical spectrum was converged with the top eight valence and lowest eight conduction bands. 

\begin{figure*}
	\centering
	\includegraphics[width =0.9\textwidth]{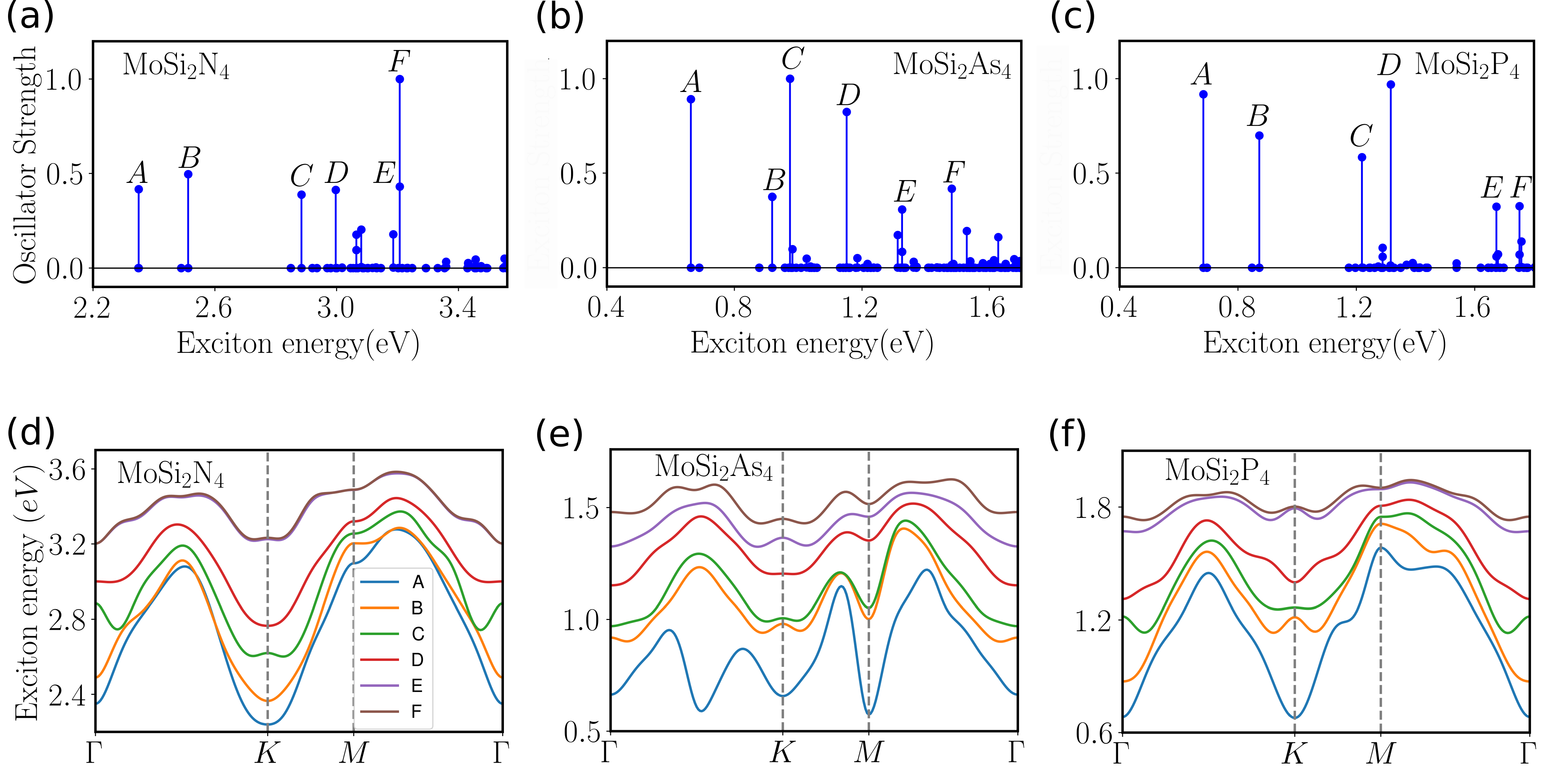}
	\caption {The exciton oscillator strength and the energy location of all the excitons in monolayers of a) \formulaN, b) \formulaAs, and c) \formulaP. Large dipole oscillator strength is generally indicative of the brightness of an exciton. Each of the three \formula~ monolayers supports at least six prominent bright excitons in the electronic bandgap region.  Finite momentum exciton bandstructure for the six bright excitons listed in Table~\ref{table:2} for the monolayers (d) {\formulaN}, (e) {\formulaAs}, and (f) {\formulaP}, respectively.}  
	\label{EE}
\end{figure*}

To probe the impact of pump fluence on the excitons binding energies, we study the non-equilibrium carrier dynamics of the system. We perform real-time simulation using the td-BSE equation~\cite{Marini} and the non-equilibrium Green's functions (NEGFs) technique as implemented in the YAMBO  code. 
The non-equilibrium population of the photo-excited electronic states in the presence of the pump laser pulse is obtained by following the time evolution of the density matrix. The equation of motion for the density matrix is projected onto 20 bands. To account for the dissipative effects in the dynamics, a relaxation term with different scattering timescale for the population relaxation and dephasing is added to the propagation equation for the Green's function [Eq. (11) of Ref.~\cite{Marini}]. We have chosen 650 $fs$ as the scattering timescale of the perturbed electronic population and 100 $fs$ for the dephasing rate. The pump field is simulated as a sinusoidal time-dependent external potential (of a specific frequency) convoluted with a Gaussian function in time. 
We have chosen the full-width at half maximum (FWHM) to be 100 $fs$. The intensity of the applied field is varied from $(6-70) \times 10^5$ ($kW/cm^2$). 

\section{EQUILIBRIUM PROPERTIES}
\label{EQP}
Experimentally synthesized monolayer {\formulaN} crystal structure consists of seven
    atomic layers in the sequence of N-Si-N-Mo-N-Si-N as shown in Fig.~\ref{str} (a)-(b). These individual atomic layers are held together by strong covalent bonds~\cite{Yu_2021}. The same structure is shared by the {\formulaAs} and {\formulaP} monolayers with different lattice parameters. For the bi-layers, the AB stacking is found to be energetically the most favourable structure~\cite{Rajibul-Barun-spin,Zhong2021}. 
To explore the light-matter interaction and optical absorption, we start from the ground state electronic bandstructure of the monolayer {\formula} series. We note that for accurate calculation of the exchange interaction in both the evGW and the BSE, it is essential to use the semi-core ($4s$ and $4p$) orbitals for the Mo atoms~\cite{spin-orbit-MoS2-PRB}.

\subsection{QUASIPARTICLE BANDSTRUCTURE}

We find that within the GGA approximation, including spin-orbit coupling (SOC), monolayer {\formulaN} is an indirect bandgap semiconductor with a bandgap of $1.78$ eV. In contrast,  the monolayers {\formulaAs}  and {\formulaP} are direct bandgap semiconductors with a bandgap of $0.51$ eV and $0.60$ eV at the $K$ point of the 2D hexagonal BZ. 
To improve the estimation of the electronic bandgap, we include the QP self-energy corrections on top of the GGA DFT theory calculations, using the GW method. Our GW calculation indicate that the monolayer {\formulaN} has an indirect bandgap of 3.58 eV, while {\formulaAs} and {\formulaP} have direct bandgap of 1.70, and 1.74 eV, respectively. Interestingly, our GW calculations for QP bandstructure indicate that the bandgap in these materials is even higher than that obtained from hybrid HSE06 functional-based calculations~\cite{Hong670,Yang2021-non-linear-optical-response}. We present the band structure of these three monolayers in Fig.~\ref{BS} (a)-(c). A comparison of the bandgap obtained from different methods is shown in Table-\ref{table:1}. 

A subtle feature of the electronic band structure is that the combined effect of the absence of inversion symmetry and the strong SOC of the Mo-$d$ orbitals breaks the valence band edge degeneracy at the $K$ point. This is similar to the valence band splitting in TMD  monolayers~\cite{SOC-effect-MoS2-PRL}. We find that the SOC splits the valence band maxima (VMB) at the K point in the GW (GGA) calculations by 154 (129), 226 (182), and 171 (138) meV in monolayer {\formulaN}, {\formulaAs}  and {\formulaP},  respectively.

\begin{table*}
\caption{The exciton energy and the dipole oscillator strength of the most prominent exciton peak (having the oscillator strength of more than 0.3 out of 1) shown in  Fig.~\ref{EE} (a)-(c) for the monolayer {\formulaN}, {\formulaAs}, and {\formulaP} respectively. The binding energies are calculated for the $A$ and $B$ excitons. In all three monolayers, there are six excitons (with multiple degeneracies shown in Fig.~\ref{EE}) found below the minimum of the non-interacting QP bandgap and are strongly bound with a BE of $~1.0$ eV or more.}
\vspace{0.15 cm}

    \begin{tabular}{c| c|c|c}
    \hline \hline
      Structure & {\formulaN} & {\formulaAs} & {\formulaP}\\
      \hline
      \begin{tabular}{c}
      Peak \\ 
     \hline
      A \\
      B\\ 
      C\\ 
      D\\
      E\\
      F\\
    \end{tabular}&\begin{tabular}{ccc}
     Energy (eV) & Strength & BE (eV)\\ 
     \hline
      2.35 & 0.42 & 1.35\\
      2.51 & 0.50 & 1.35\\ 
      2.88 & 0.39 & \\ 
      2.99 & 0.41 & \\
      3.21 & 0.43 & \\ 
      3.21 & 1.00 & \\ 
    \end{tabular} & \begin{tabular}{ccc}
      Energy (eV) & Strength & BE (eV)\\ 
      \hline
      0.66 & 0.89 & 1.04\\
      0.92 & 0.37 & 1.01\\ 
      0.97 & 1.00 & \\ 
      1.15 & 0.82 & \\
      1.33 & 0.31 & \\ 
      1.48 & 0.42 & \\ 
    \end{tabular} & \begin{tabular}{ccc c}
      Energy (eV) & Strength & BE (eV)\\ 
      \hline
      0.68 & 0.91 & 1.11\\
      0.87 & 0.70 & 1.01\\ 
      1.22 & 0.58 & \\ 
      1.32 & 0.97 & \\
      1.67 & 0.32 & \\ 
      1.74 & 0.69 & \\ 
    \end{tabular} \\ 
      \hline \hline
    \end{tabular}
    \label{table:2}
\end{table*}


Having obtained the electronic spectrum, we now focus on the optical absorption spectrum. Owing to the reduced screening of the Coulomb interactions in 2D materials, these synthetic monolayers have enhanced electron-hole interaction (EHI). This can give rise to several excitonic peaks below the electronic bandgap, similar to that found in MoS$_2$ monolayers. Thus, it is essential to consider the attraction between the quasi-electrons and quasi-holes by solving the BSE to obtain a reasonably good optical absorption spectra~\cite{strain-exciton-Mos2-PRB}. 






\subsection{OPTICAL ABSORPTION SPECTRA AND EXCITONS}

To include the Coulomb interactions between the electrons and holes in our calculations, we use the quantum MBPT (an effective two-body approach). This goes beyond the one-particle picture of individual quasi-electron and quasi-hole excitations~\cite{BSE-1-Strinati,BSE-2-Strinati,DFT_GWA_Exciton-GW3}. The EHI includes the excitonic resonances, and it drastically modifies the corresponding single-particle optical absorption spectra~\cite{Hanke1979-PRL}. Specifically, we solve the Bethe-Salpeter equation of motion for two-particle Green's
function, starting from the single-particle Green's function. We start with the QP electron and hole states and their QP energies and calculate the impact of the EHI via the self-energy terms in the electron-hole propagator. The solution of the BSE yields the coupled and correlated electron-hole excitation states~\cite{BSE-1-Strinati,Hanke1979-PRL,SOC-effect-MoS2-PRL,Mishra-2020-PRB}. 

The optical absorption spectra is defined by the imaginary part of the dielectric function $\epsilon (E)$, which is specified by 
\begin{equation}
    \epsilon_2(E) \propto \sum_\lambda |\textit{\textbf{D}}|^2\delta(E^\lambda -E).
    \label{eq:abs}
\end{equation}
Here, $E^\lambda$ are the exciton energy eigenvalues calculated from the BSE for the exciton state $\lambda$. The dipole oscillator strength (OS) is defined as 
\begin{equation}
\textit{\textbf{D}} = \sum_{cv}{A_{cv}^\lambda} \langle c|\textbf{d}|v\rangle~.
\label{os}
\end{equation}
The dipole matrix element $\langle c|\textbf{d}|v\rangle$ captures the electronic transitions from the valence ($v$) to the conduction ($c$) band  states~\cite{yambo2019,spin-orbit-MoS2-PRB}. $A_{cv}^\lambda$ are the expansion coefficients of the exciton state $\lambda$ in the electron-hole basis. 

\begin{figure*}[t]
	\centering
	\includegraphics[width =0.75\textwidth]{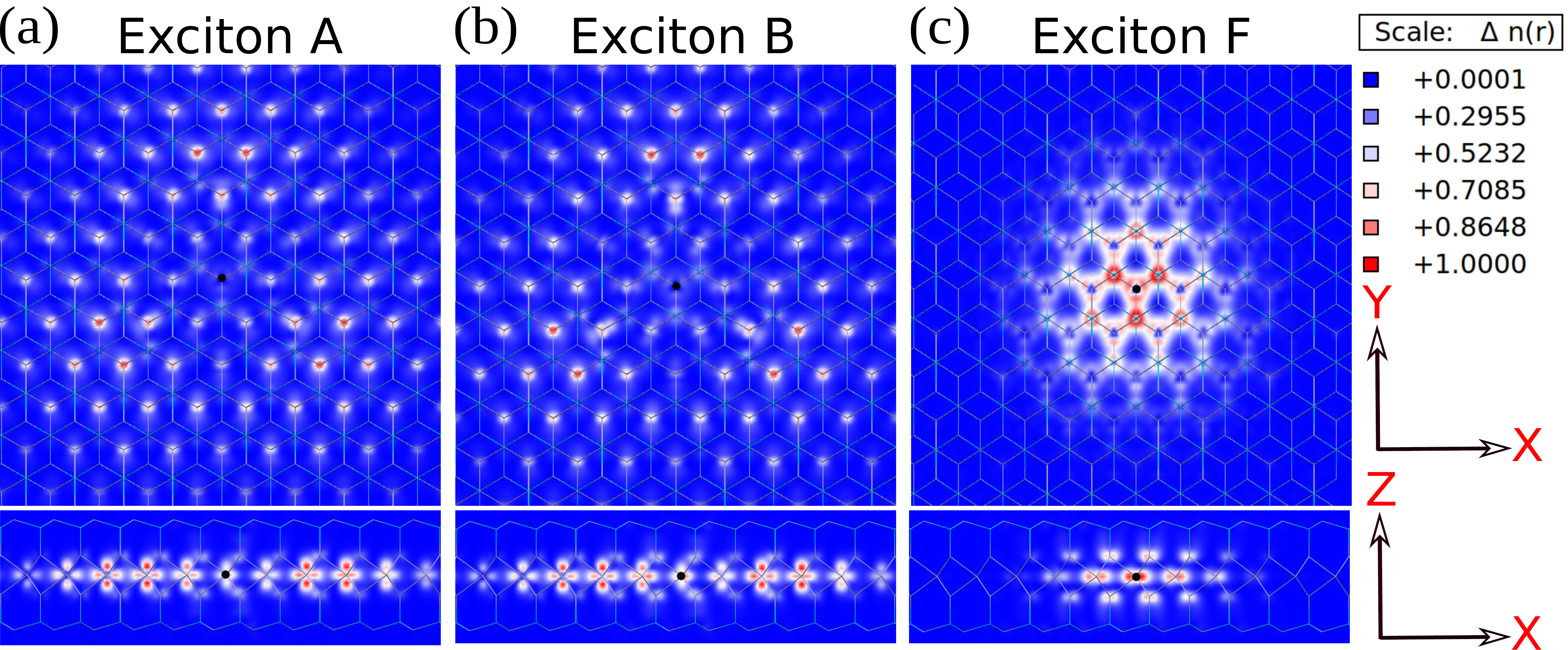}
	\caption {The real space probability density of the {\formulaN} (a) A exciton, (b) B exciton, and (c) F exciton wavefunction. In all figures, i) the top panel shows the 2D $x-y$ plane, ii) the bottom panel shows the out-of-plane ($x-z$)  view, and iii) the hole has been placed on the Mo atom (black dot) in the center. The A and the B exciton have a similar probability distribution, indicating that they have a similar origin. Analyzing the momentum resolved oscillator strength for these excitons confirms that they arise from the direct transitions from the spin split valence bands at the $K$ and $K'$ point of the BZ. 
	}
	\label{Exciton}
\end{figure*}

We present the calculated optical absorption spectrum for monolayer {\formulaN}, {\formulaAs}, and {\formulaP}, including the EHI (blue color) and excluding the EHI (orange color), in Fig.~\ref{BS} (d)-(f). The calculated optical bandgap (presented in the Table-~\ref{table:1}) is in good agreement with the previously reported theoretical~\cite{exciton-1-wu2021mosi2n4,Bafekry2020MoSi2N4SA} and experimental~\cite{Hong670} values. 
Clearly, the inclusion of excitonic effects changes the optical spectrum significantly and reduces the optical bandgap by almost 1 eV in all three {\formula} monolayers. 
Additionally, the excitonic absorption spectrum shows multiple prominent excitonic peaks even in the QP bandgap region. 
We have explicitly marked the location of the first three bright exciton peaks as $A$, $B$, and $C$ exciton peaks lying in the QP bandgap region in Fig.~\ref{BS} (d)-(f). 

To identify all the excitonic states in the three \formula~ monolayers, we analyze the positive eigenvalues solutions of the exciton eigenvalue equation. It is given by~\cite{DFT_GWA_Exciton-GW3},  
\be
(E_c - E_v)A^{\lambda}_{cv} + \sum_{v'c'}K^{AA}_{vc,v'c'}(E_{\lambda})A^{\lambda}_{v'c'} = E^{\lambda}A^{\lambda}_{vc}~.
\ee
Here, $E_c$ and $E_v$ are the QP energies of the electron and hole, respectively. The $K^{AA}_{vc,v'c'}(E_{\lambda})$ represent the matrix elements of the EHI kernel. 
The energy location of all the different exciton states, and their oscillator strength, are presented in Fig.~\ref{EE} (a)-(c), for all the three {\formula} monolayers. We find several bright exciton peaks in the QP bandgap region of these monolayers. Six  prominent bright excitons, with the largest oscillator strength are explicitly marked by alphabetic letters $A-F$ in Fig.~\ref{EE} (a)-(c). The properties of these six prominent bright excitons are summarized in  Table~\ref{table:2}. Specifically, we find that the two of the lowest energy excitons, the $A$ and $B$ exciton peaks, in all three monolayers are doubly degenerate. 
Each of these doubly degenerate excitons consists of a bright exciton (with a large oscillator strength) and a dark exciton (with a vanishingly small oscillator strength), as shown in Fig.~\ref{EE} (a)-(c). The excitonic peaks $A$ and $B$ correspond to electron-hole pairs arising from the direct transition from the SOC split valence bands to the conduction band at the $K$ point of the BZ. Both of these features are very similar to the $A$ and $B$ exciton peaks reported in monolayer MoS$_2$~\cite{optical-MoS2-PRL,spin-orbit-MoS2-PRB}. We find that both the $A$ and $B$ excitons are strongly bound and have a very large BE $\sim 1.35$ eV for monolayer {\formulaN}. To the best of our knowledge, this is the largest exciton BE reported for any 2D material. For example, the BE of the most tightly bound exciton~\cite{optical-MoS2-PRL,TMD-BE} in MoS$_2$, WS$_2$ and in WSe$_2$ is 0.96 eV, 0.83 eV, and 0.79 eV, respectively. {\formulaAs} and {\formulaP} also display similar excitonic peaks in their absorption spectrum shown in Fig.~\ref{BS} (d)-(f). The $A$ and $B$ exciton of the monolayer {\formulaAs} and {\formulaP} also have a very large BE of around 1 eV. 

Finite-momentum excitons are optically dark, but they play an important role in hot carrier relaxation and valley dynamics~\cite{Wang2013,Exciton_band_mos2_PRB}. The finite momentum exciton bandstructure can be probed via momentum-resolved electron energy loss spectroscopy and via nonresonant inelastic x-ray scattering~\cite{exciton_BS_C3N}. To calculate the exciton bandstructure, we solve  the BSE for finite momentum at the irreducible momentum points of the BZ. Using this, we interpolate the exciton bandstructure along the high-symmetry direction in the BZ of the {\formula} monolayer. 
We find that several excitons (degenerate dark and bright excitons at the Gamma point or $q=0$) preserve their degeneracy at the $K$ and $M$ high-symmetry points. However, these degenerate exciton bands split at some generic $q$ points in the BZ. We present the exciton dispersion of the six prominent excitons listed in Table~\ref{EE}, for monolayer {\formulaN}, {\formulaAs}, and {\formulaP} in Fig.~\ref{EE} (d)-(f). We find that the lowest energy excitons have almost equal energies at the $\Gamma$ and the $K$ points, similar to that found in MoS$_2$~\cite{Exciton_band_mos2_PRB} and hexagonal boron nitride~\cite{exciton_BS_hBN_PRB}. The lowest bright excitons ($A$) has an almost linear dispersion around the $\Gamma$ point, similar to that observed in other 2D materials~\cite{exciton_BS_PRM_Daniele} but a quadratic behavior around the $K$ point in {\formula} monolayers~\cite{exciton_BS_PRL}. 

\begin{figure*}[!t]
	\centering
	\includegraphics[width =0.7\textwidth]{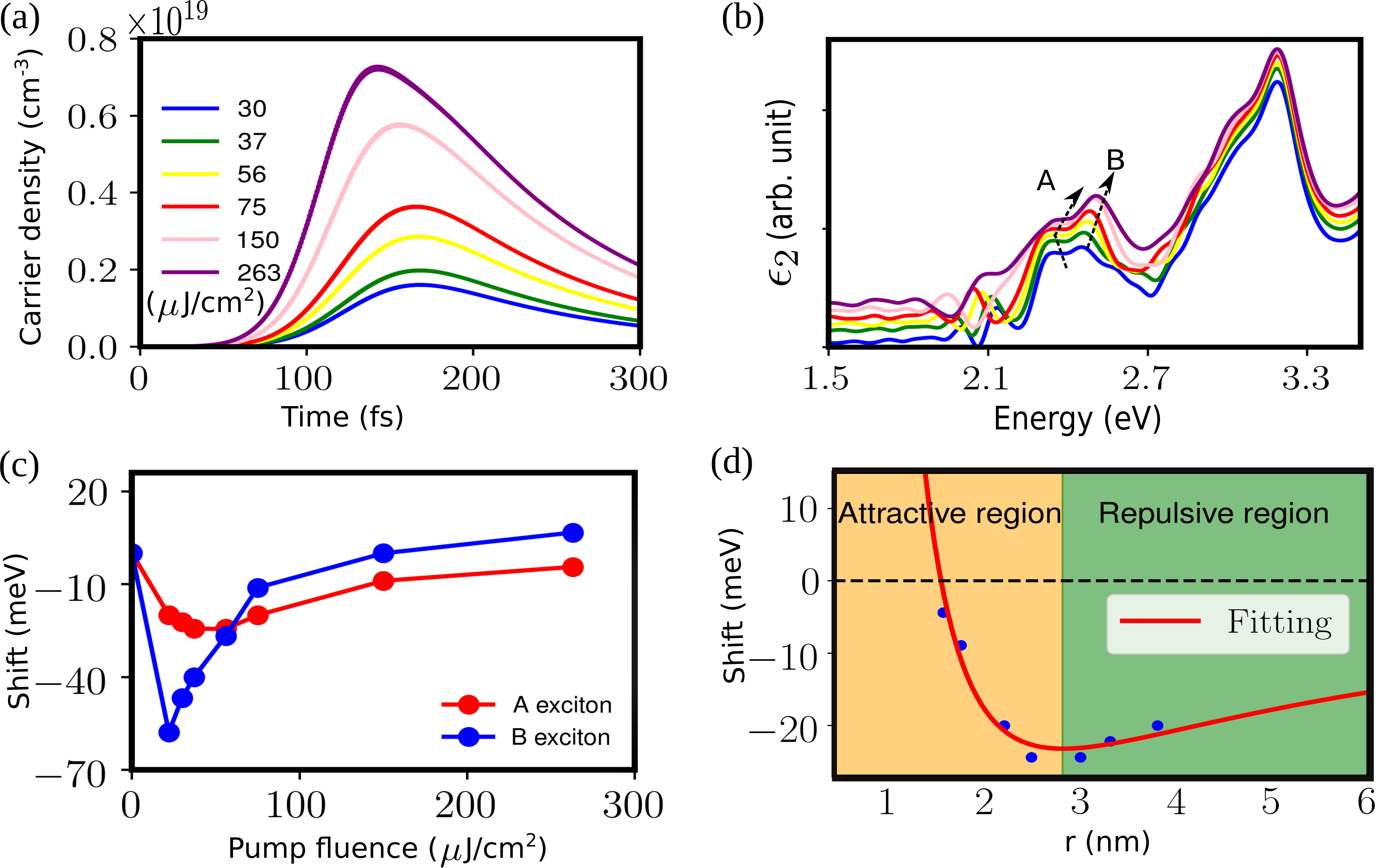}
	\caption {(a) The evolution of photo-excited carriers in \formulaN~with time, for different values of pump fluence ranging from 30-263 $\mu$J/cm$^2$, for a gaussian pump beam with full width at half maximum being 100 $fs$.  The excitation frequency of the pump beam is set at the energy of the first excitonic peak ($\omega_{pump}$= 2.35 eV) in \formulaN.
	 (b) The pump-fluence dependent optical absorption spectrum, with the $A$ and the $B$ exciton peaks marked. The color of the curves represents a different value of the pump fluence, as indicated in (a).
	(c) Pump-fluence induced shift in the binding energies of the $A$ and $B$ exciton peaks. The redshift-blueshift crossover in the BE of both the $A$ and the $B$ exciton with increasing pump fluence can be clearly seen. (d) The redshift-blueshift crossover in the BE highlights that the exciton-exciton interactions mimic the atom-atom interactions captured by a Lennard-Johnes like potential. Here, the blue dots capture the calculated BE shift for the $A$ exciton in c), while the red line is the fit to a Lennard-Johnes like potential of Eq.~\eqref{LJ}. The fitted values  of $r_0$, $p$ and $q$ are 1.53 nm, 2.3, and 1.13, respectively. The inter-exciton separation $r$ is obtained from the fluence dependent density of the photo-excited carrier density $n$, via  the relation  $\pi r^2  n = 1$.} 
	\label{TD}
\end{figure*}

To visualize the spatial distribution of the excitons for {\formulaN}, we plot the exciton probability density for the $A$, $B$, and $F$ exciton peaks in Fig.~\ref{Exciton}. We fix the hole position at the top of a Mo atom at the center in each panel (black dot). 
We find that the exciton wavefunctions of the $A$ and $B$ excitons are almost identical, suggesting a similar origin for both of these. This is also expected from the fact that the $A$ and $B$ exciton peaks arise from direct transitions from the spin split valence bands at the $K/K'$ point of the BZ. 
The exciton wavefunction of the $A$ and $B$ exciton peaks are spread over 4-5 unit cell in the real lattice structure, suggesting that these excitons are of Wannier–Mott type ~\cite{optical-MoS2-PRL,exciton_wf_ws2}. In contrast, the real space wavefunction of the brightest exciton peak $F$ shown in Fig.~\ref{Exciton} (c), has a relatively localized wavefunction. The analysis of the momentum resolved exciton oscillator strength shows that the $F$ exciton originates from  
several direct band transitions around the minimum QP bandgap, from different k-points in the BZ~\cite{exciton-1-wu2021mosi2n4,optical-MoS2-PRL}. 
All three excitons in Fig.~\ref{Exciton} show strong in-plane confinement of the exciton wavefunction, similar to that observed in other 2D materials.  This 2D confinement indicates reduced dielectric screening in the out-of-plane direction 
~\cite{hBN_excitonWF,hBN_excitonWF2}.
Having explored the excitonic equilibrium properties of the monolayer MoSi$_2$Z$_4$ series, we now focus on the renormalization of the exciton binding energies induced by increasing the pump fluence. 

\section{Carrier dynamics and its effect on absorption spectrum}
\label{Non-EQL}

To study the non-equilibrium optical properties of monolayer {\formula} series, we use the td-BSE framework as implemented in the YAMBO code~\cite{Roth,Marini,Pedio}.  
In our real-time simulation, we apply a pump electric field with frequency locked to equilibrium location of the $A$ exciton peak in \formulaN. The photo-excitation generates free electrons and holes. The time evolution of the photo-excited charge carriers for different values of the pump fluence is shown in Fig.~\ref{TD} (a). More pump fluence 
generates more photo-excited carriers over the entire pulse duration. For a particular value of the pump fluence, the number of carriers first increases with time, and then it decreases as expected for a gaussian pump pulse. The variation of the non-equilibrium absorption spectrum for different values of the pump fluence is shown in Fig.~\ref{TD} (b).  
The changing location of the exciton peaks highlights the impact of the Coulomb screening induced by the photo-excited charge carriers. 

We present the renormalization of the BE of the $A$ and $B$ exciton peaks with increasing pump fluence in Fig.~\ref{TD} (c). The exciton BE first shows a redshift with increasing pump fluence. This decrease in the BE arises from the screening of the excitonic Coulomb potential induced by the photo-excited charge carriers. However, beyond a certain density of photo-excited carriers and excitons, the exciton BE starts increasing. In this regime, the exciton-exciton repulsion starts playing a dominant role. A similar redshift to blueshift crossover of the exciton BE with increasing pump fluence obtained here for MoSi$_2$N$_4$, has been recently demonstrated experimentally for MoS$_2$~\cite{amit-exciton-2}, and also for WS$_2$~\cite{Sei}. 

The redshift to blueshift crossover in the exciton BE is similar to the atom-atom interaction with changing inter-atomic separation. To show this explicitly, we fit the exciton BE to the inter-atomic interaction potential energy specified by
\begin{equation}
\delta E =A\left[\left(\frac{r_{0}}{r}\right)^{p}-\left(\frac{r_{0}}{r}\right)^{q}\right]~.
\label{LJ}
\end{equation}
Here, $r$ is the separation between the interacting atoms or excitons. In Eq.~\eqref{LJ} $r_0$, $p$, $q$ and $A$ are treated as fitting parameters. By assuming that all the photo-excited carriers form excitons, $r$ can be related to the photo-excited carrier density ($n$) via the condition, $\pi r^{2} n = 1$. Fitting yields, $r_0 = 1.53$ $nm$, $p = 2.30$ and $q = 1.13$. The minima of the exciton-exciton interaction potential occurs at, $r_{\rm eq} = 2.9$ $nm$.  
We show in Fig.~\ref{TD} (d) that the exciton BE fits well to the interaction potential specified by Eq.~\eqref{LJ}. 

\section{Summary}
\label{SC}
We have demonstrated that the {\formula} series of monolayers hosts very strongly bound excitons, with potential for optoelectronic applications in the infrared and in the visible regime. By explicitly including the QP self-energy corrections in GW based bandstructure calculations, we find that monolayer MoSi$_2$N$_4$ 
hosts an indirect bandgap of 3.58 eV with a comparable 
direct bandgap. In contrast, the monolayer MoSi$_2$As$_4$ and MoSi$_2$P$_4$ have a direct bandgap of 1.70 eV and 1.74 eV, respectively. Starting from the GW based QP bandstructure calculations and including the two-particle electron-hole correlations, we show the existence of several (around six) strongly bound bright excitons within the QP bandgap region in this series of materials. The binding energies of the prominent $A$ and $B$ exciton peaks in all three monolayers are greater than 1 eV, and it can be as large as 1.35 eV in {\formulaN}.
Additionally, we solve the td-BSE equation to study the fluence-dependent renormalization of the excitonic BE. Interestingly, we find a redshift-blueshift crossover in the exciton BE with increasing pump fluence, indicating atom-like interactions between excitons with increasing exciton density. Our study establishes the monolayers of synthetic {\formula} series to be an exciting platform for exploring the physics of strongly bound excitons and their non-equilibrium dynamics. 

\section{ACKNOWLEDGEMENTS}
We thank Dr. Barun Ghosh and Dr. Sitangshu Bhattacharya for very helpful discussions. The work at TIFR Mumbai was supported by the Department of Atomic Energy of the Government of India under Project No. 12-R$\&$D-TFR-5.10-0100.
We acknowledge the Science and Engineering Research Board (SERB) and the Department of Science and Technology (DST) of the Government of India for financial support. We also acknowledge the high-performance computing facility at IIT Kanpur for computational support.
We acknowledge the National Supercomputing Mission (NSM) for providing computing resources of `PARAM Sanganak' at IIT Kanpur, which is implemented by C-DAC and supported by the Ministry of Electronics and Information Technology (MeitY) and Department of Science and Technology (DST), Government of India. P. Y. acknowledges the UGC for Senior Research Fellowship. B. K. acknowledges DAE (Govt. of India) via sanction no. 58/20/15/2019-BRNS for the research fellowship. 

\bibliography{myref}
\end{document}